\documentclass[10pt,sort&compress,3p]{elsarticle}
\usepackage{amsmath,amssymb,amsthm,mathrsfs,amsopn}
\usepackage{graphicx}
\usepackage{subfigure}
\usepackage{booktabs}
\usepackage{array}
\usepackage{tcolorbox} 
\tcbset{colback=blue!5!white,colframe=blue!50!black, left=0mm,right=0mm,top=0mm,bottom=0mm, box align=center, halign=center} 
\usepackage{tikz} 
\usetikzlibrary{automata}
\usepackage{enumitem}
\usepackage{fancyhdr} 
\usepackage{pdfpages} 
\usepackage{longtable}

\journal{Communications in Nonlinear Science and Numerical Simulation}

\begin{document}

\begin{frontmatter}

\title{Iterative procedure for network inference}

\author[1]{Gloria Cecchini\corref{cor1}}\ead{$gloria.cecchini@unifi.it$}
\author[2]{Bj\"orn Schelter}\ead{$b.schelter@abdn.ac.uk$}
\cortext[cor1]{Corresponding author}
\address[1]{CSDC, Department of Physics, University of Florence, via G. Sansone 1, 50019, Sesto Fiorentino, Florence, Italy}
\address[2]{Institute for Complex Systems and Mathematical Biology, University of Aberdeen, Meston Building, Meston Walk, Aberdeen, AB24 3UE, United Kingdom}

\date{\today}

\begin{abstract}
When a network is reconstructed from data, two types of errors can occur: false positive and false negative errors about the presence or absence of links.
In this paper, the vertex degree distribution of the true underlying network is analytically reconstructed using an iterative procedure.
Such procedure is based on the inferred network and estimates for the probabilities $\alpha$ and $\beta$ of \textit{type I} and \textit{type II errors}, respectively.
The iteration procedure consists of choosing various values for $\alpha$ to perform the iteration steps of the network reconstruction.
For the first step, the standard value for $\alpha$ of $0.05$ can be chosen as an example.
The result of this first step gives a first estimate of the network topology of interest.
For the second iteration step the value for $\alpha$ is adjusted according to the findings of the first step. 
This procedure is iterated, ultimately leading to a reconstruction of the vertex degree distribution tailored to its previously unknown network topology.
\end{abstract}

\begin{keyword}

network inference \sep node degree distribution  \sep false positive, false negative \sep statistical inference

\end{keyword}

\end{frontmatter}

\section{Introduction}

Complex systems are of key interest in multiple scientific fields, ranging from medicine via physics, mathematics, engineering to economics \citep{Barrat2008,Boccaletti2006,Cohen2010}.
These systems can be modelled, or represented as networks, where vertices are the elements of the system and links represent the interactions between them. 

Networks are ubiquitous in many fields of study \citep{Fagiolo2007}; examples include the Internet, airline connections, scientific collaborations and citations, trade market contacts, social relationships, cellular and ecological systems, transportation systems, power grids, and the human brain \citep{Barabasi2002, Banavar, Bar2003, Dorogovtsev2003, Egu, Kurant, Newman2003, Odor, Valencia, Xue, Yeung}. 

In a more theoretical framework in physics, networks are also studied to investigate synchronization phenomena of coupled oscillators as well as the analysis of chaotic behaviour and corresponding phenomena in dynamical systems \citep{Rok2017,Arkady,Li2014,Arkady2016}.

To capture particular features of a network, properties or characteristics have been introduced in the literature \citep{Olbrich2010}.
For instance, the vertex degree describes the number of links of a vertex \citep{NewmanBook}.
The vertex degree distribution is a key property to study the general structure of the connectivity of a network, and it is predominantly used in this paper for this reason.

Some complex systems can be directly observed such as power grids, or transportation systems.
In these kinds of systems, the network topology is known a priori, and its characteristics can be investigated.
Other complex systems cannot be directly observed, such as the functional network of the human brain, therefore network structure is not known a priori, and must be inferred indirectly.
When the underlying network is not known a priori, reliably inferring network structure from data is crucial to represent the system accurately; this is known as the inverse problem \citep{Asllani2018,Burioni2014,Shandilya2011}.

When a network is to be inferred from data, typical analysis techniques provide a measure of connectivity strength for each link.
Statistical methods are then used to decide whether these measures pass a certain threshold, and thereby provide a means to decide if the corresponding links are considered present.
If a link is erroneously detected, this is called false positive link and it is referred to as a \textit{type I error}.
Likewise, an existing link that remains undetected is called false negative link and it is referred to as a \textit{type II error}.
The probability of detecting a false positive link is usually denoted by $\alpha$, while $\beta$ refers to the probability that an existing link remains undetected.

A prototypical class of systems that cannot be directly observed consists of dynamical systems where only the dynamic of the nodes is accessible.
In this case, various strategies have been suggested in the literature to reconstruct the network structure \citep{Rok2017, Arkady, Li2014, Arkady2016, Asllani2018, Burioni2014, Shandilya2011, Leguia2019, Banerjee2019, Panaggio2019}. 
For the sake of simplicity, the reader can consider this kind of systems to be reconstructed; nevertheless, the method described in this paper is more general, and it can be applied once a measure of connectivity strength for each link is established.

Classical statistical methods aim to reconstruct with high certainty the presence of links, i.e., the analysis has high specificity, and the standard value of 0.05 for $\alpha$, i.e. the significance level of the test, is often chosen \citep{Chavez2010, Fallani2014, devore2011,Honey2007, jalili2011, quinn2002,Schinkel2011}.
The decision to set $\alpha=0.05$ does not take into account the probability of false negative links, since, usually, high specificity implies low sensitivity, meaning a high chance of missing links.
Intuitively, there is an inverse relationship between the probabilities of \textit{type I} and \textit{type II errors}; hence, it is typically impossible to have both $\alpha$ and $\beta$ equal to zero.

In~\citep{GloriaJNM2018} the authors describe the influence of false positive and false negative conclusions about links on the network structure.
Several simulation results are presented and optimal values for each network topology and characteristic are shown.
In the conclusions, the authors speculate that the simulation study presented could be used as an iterative procedure to achieve a better network reconstruction.
This manuscript is dedicated to such a procedure.
The results obtained in~\citep{GloriaPRE2018} are implemented in this procedure, since they provide an analytic framework to reconstruct the vertex degree distribution of the network.

Roughly, the iteration procedure consists of choosing various values for $\alpha$ to perform the iteration steps of the network reconstruction.
For the first step, the standard value for $\alpha$ of $0.05$ can be chosen as an example.
The result of this first step gives a first estimate of the network topology of interest.
For the second iteration step the value for $\alpha$ is adjusted according to the findings of the first step. 
This procedure is iterated, ultimately leading to a reconstruction of the network characteristic tailored to its previously unknown network topology.

This paper is structured as follows. 
Section \ref{preliminar} is dedicated to present some preliminary results needed for the implementation of the iterative procedure presented in this paper.
In Sec. \ref{Procedure} such iterative procedure for network inference is presented.
Section \ref{ResultsProcedure} shows a simulation study of the method depicted in Section \ref{Procedure}.

\section{Preliminary results}
\label{preliminar}

In this section we give some preliminary results that will be needed to implement the iterative procedure presented later in Sec.~\ref{Procedure}.
In particular, Sec.~\ref{preliminary1} introduces some concepts of test of hypothesis and Sec.~\ref{preliminary2} is dedicated to summarise the results obtained in \citep{GloriaPRE2018}.

\subsection{Type I and type II errors}
\label{preliminary1}

In \citep{GloriaJNM2018}, it is shown that $\alpha$ and $\beta$ are reciprocally dependent.
Their functional relationship depends on the nature of the problem taken into account.
Consider, for example, a network of coupled oscillators, and assume to be able to detect the dynamic of every vertex.
Various strategies have been suggested in the literature to recover a measure for the strength of connection for every link \citep{Rok2017,Arkady,Li2014,Arkady2016}.
For instance, correlation can be used as a measure to this aim, and it will be employed throughout this manuscript.
Therefore, assume to assign a correlation value to each couple of nodes, and appoint it to the respective link.

When the correlation coefficients are estimated from a sample, they have a certain distribution that depends on the dimension of the sample and the true correlation coefficient.
As shown in \citep{kenney1951}, for data that follow a bivariate normal distribution, the exact probability density function of the estimated correlation coefficients $r$ for a sample of $N$ data points with true correlation coefficient $\rho$ is
\begin{equation}
f(N,\rho,r)=\frac{(N  -  2)\Gamma(N  -  1)(1  - \rho^2)^\frac{N - 1}{2}(1  - r^2)^\frac{N - 4}{2}}{\sqrt{2\pi}\ \Gamma(N  - \frac{1}{2})(1  - r\rho)^{N - \frac{3}{2}}}\quad {}_2F_1\left(\frac{1}{2},\frac{1}{2};\frac{2N  -  1}{2};\frac{r\rho + 1}{2}\right),
\label{correlation}
\end{equation}
where $\Gamma$ is the gamma function and ${}_2F_1(\cdot,\cdot;\cdot;\cdot)$ is the Gaussian hypergeometric function.

When performing a hypothesis test for correlation for every pair of vertices, a $p-$value of the test is found.
If the $p-$value is smaller than or equal to the significance level $\alpha$, then the link between the corresponding vertices is considered to be present.
Call $r_\tau$ the value of the correlation $r$ such that 
\begin{equation}
\alpha=\int_{r_\tau}^1 f(N,0,r) \ dr,
\label{eqAlpha}
\end{equation}
meaning that the rejection region is obtained for coefficients $r_\tau\leq r\leq 1$.
Consequently, estimated coefficients $r<r_\tau$ do not lead to a rejection of the null hypothesis.
The probability of false negative conclusions
\begin{equation}
\beta=\int_{-1}^{r_\tau} f(N,\rho,r) \ dr
\label{eqBeta}
\end{equation}
is calculated once $r_\tau$ is fixed.

Once $\alpha$ is chosen,
\begin{equation}
\beta=\beta(N,\rho,\alpha)
\label{betaAlphaNrho}
\end{equation}
can be found as a function of $\alpha,N,\text{ and }\rho$.

\begin{figure}[t]
\centering
\includegraphics[scale=0.48]{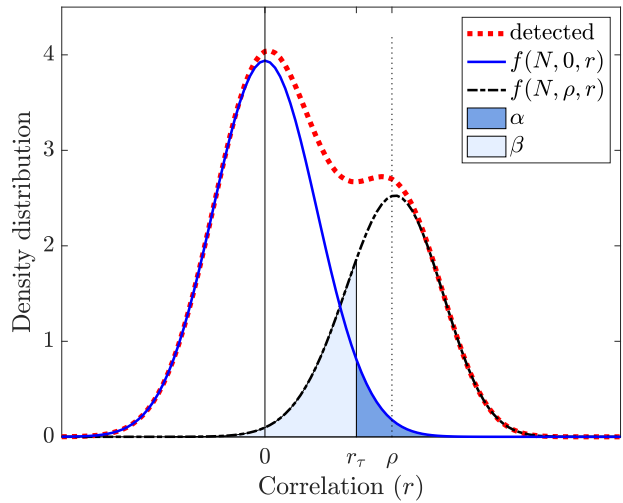}
\caption{Density distribution of correlation coefficients $r$ for a sample of $N=100$ data points with true correlation coefficient $\rho=0$ (solid blue line) and $\rho=0.25$ (dashed black line), i.e. $f(100,0,r)$ and $f(100,0.25,r)$ as described by Eq. (\ref{correlation}). The aggregated detected distribution (red dotted line) is given by the weighted sum $(1-p^*) f(N,0,r)+p^* f(N,\rho,r)$. The area in blue is the $\alpha$ value corresponding to $r_\tau=0.18$, and the area in light blue is the respective value of $\beta$.}
\label{fig_alphabeta}
\end{figure}

In the case of network reconstructions, after the value of correlation is estimated for each link, a threshold $r_\tau$ is chosen, and $\alpha$ and $\beta$ are inferred.
This is equivalent to infer $f(N,0,r)$ and $f(N,\rho,r)$ from the aggregated distribution of all the estimated correlation values.
The detected distribution of all the estimated correlation values is indeed the weighted sum of $f(N,0,r)$ and $f(N,\rho,r)$, for the absent and true links in the underlying network.
As an example, Fig.~\ref{fig_alphabeta} shows the detected distribution in red dotted line, the distributions $f(N,0,r)$ and $f(N,\rho,r)$ in solid blue and dashed black lines, respectively.
The dark and light blue areas correspond to $\alpha$ and $\beta$, respectively, for the fixed value of $r_\tau$.
Given the rate $p^*$ of true links with respect to the total, the detected distribution is given by $(1-p^*) f(N,0,r)+p^* f(N,\rho,r)$.

In a real-world application, an estimate for the distribution of the correlation coefficients as a function of the number of data points $N$ and the true correlation coefficient $\rho$ should be available.
Nevertheless, there exist cases in which $\rho$ is not the same for every link; consider, e.g., a network of coupled oscillators where the initial coupling strengths are not the same for all the links.
For sake of simplicity, and to make the argument clearer, this manuscript assumes that these correlation coefficients are independent and identically distributed (iid) and follow the distribution $f(N,\rho,r)$ in Eq.~(\ref{correlation}).
In this case, it is reasonable to assume that it is possible to estimate $\rho$, given the distribution of the correlation coefficients.

A more complicated scenarios consists of correlation coefficients which are still independent, but follow more than one distributions $f(N,\rho,r)$, i.e., there is not a unique value for $\rho$.
In this case, it still should be possible to estimate the distributions $f(N,\rho,r)$, and therefore apply the method developed in this paper, using different values for $\beta$, as shown in Eq.~\ref{eqBeta}.
A generalization for correlation coefficients which are not independent is likely more complicated, and therefore requires a more in-depth analysis.

\subsection{Analytical framework on network inference}
\label{preliminary2}

The work presented in~\citep{GloriaPRE2018} shows that the vertex degree distribution of a network is influenced by false positive and false negative conclusions about the presence or absence of links.
Here we summarise the main results that will be used later for the iterative procedure.

Consider a network $G$ with $n$ vertices and vertex degree distribution defined by the probability function $\mathcal{P}$.
Call $G'$ the network detected when \textit{type I} and \textit{type II errors} occur; as mentioned above, $\alpha$ is the probability of a \textit{type I error} and $\beta$ is the probability of a \textit{type II error}.
Hence, the set of edges of $G'$ is a combination of true positive links and false positive links of $G$.
The vertex degree distribution of $G'$ is characterised by the probability function $\mathcal{P'}$.

The probability that a vertex has degree $k'$ in $G'$, knowing it has degree $k$ in $G$ is 
\begin{equation}
\mathbb{P}(d'=k'|d=k)=\sum_{i=\max\{0,k-k'\}}^{\min\{k,n-1-k'\}}\binom{k}{i}(1-\beta)^{k-i}\beta^i\binom{n-1-k}{k'-k+i} \alpha^{k'-k+i}(1-\alpha)^{n-1-k'-i},
\label{compactEq}
\end{equation}
which corresponds to Eq.~(A1) in~\citep{GloriaPRE2018}.
Hence, calling $\mathcal{P}_i=\mathbb{P}(d=i)$ and $\mathcal{P}'_i=\mathbb{P}(d'=i)$, Eq.~(\ref{compactEq}) can be written in a matrix form as 
\begin{equation}
\label{p1Ap}
\mathcal{P'}=A\mathcal{P},
\end{equation}
where $A$ is the matrix with elements the conditional probabilities $A_{k+1,k'+1}=\mathbb{P}(d'=k'|d=k)$; note that $k,k'\in\{0,\cdots ,n-1\}$. 
Equation~(\ref{p1Ap}) corresponds to Eq.~(7) in~\citep{GloriaPRE2018}.
The matrix $A=A(n,\alpha,\beta)$ depends on $n$, $\alpha$ and $\beta$ and has determinant $\det A=(1-\alpha-\beta)^{\frac{n(n-1)}{2}}$, then it is invertible if and only if $\alpha\neq 1-\beta$.
Note that, since the convergence to zero of the determinant of $A$ scales like $x^{\frac{n(n-1)}{2}}$ for $|x|<1$, numerical issues arise for relatively small $n$ when inverting the matrix $A$ to find $\mathcal{P}$ through $\mathcal{P}=A^{-1}\mathcal{P'}$.

The vertex degree distribution of the original network is found carefully inverting Eq.~(\ref{p1Ap}).
Call $A_t^+$ the pseudoinverse of the truncated matrix $A$ using the singular value decomposition, i.e., only the first $t$ singular values are considered different from zero.
A method to assess optimal value for $t$ has been studied in \citep{Frank,Gavish2014}, but it still remains an open problem.
In this paper, the choice for $t$ lies on the difference between consecutive singular values, i.e., $t$ is the largest index for the singular values $\sigma_i$ such that $\sigma_t-\sigma_{t+1}\geq\sigma_i-\sigma_{i+1}$ for all $i$.
The vertex degree distribution of the original network
\begin{equation}
\mathcal{P}=A_t^+\ \mathcal{P'}
\label{pA+p1}
\end{equation}
is thus calculated analytically if the biased one and the probabilities of \textit{type I} and \textit{type II errors} are given.
Note that the matrix $A$ depends on $n,\alpha,\beta$, where $n$ is the number of vertices in the network.

As stated above, this reconstruction method assumes the probabilities of \textit{type I} and \textit{type II errors} to be known a priori.
To show the impact of violations of this and thereby the robustness of the method, the performance of the reconstruction is analysed when perturbations on $\alpha$ and $\beta$ are introduced.
The perturbations are quantified in percentage using the parameter $\delta$, e.g. the perturbations of $\beta$ are expressed by $\beta^p=\beta+\delta\beta$.
Negative values for $\delta$ represent underestimated values for $\beta$ and positive overestimated values for $\beta$.
The same argument is used for the perturbation of $\alpha$.

The robustness analysis shows that a rough estimate for $\alpha$ and $\beta$ is sufficient to get an accurate reconstruction; the method is robust to relatively large perturbations of these two errors. 
In~\citep{GloriaPRE2018} various example are presented to demonstrate the robustness of the approach.
Here we present one of those cases to briefly explain the robustness analysis which will be used later for the iterative procedure in Sec.~\ref{Procedure}.

\begin{figure}[!t]
\centering
\includegraphics[width=0.5\textwidth,trim=0.9cm 0cm 1cm 0cm,clip=true]{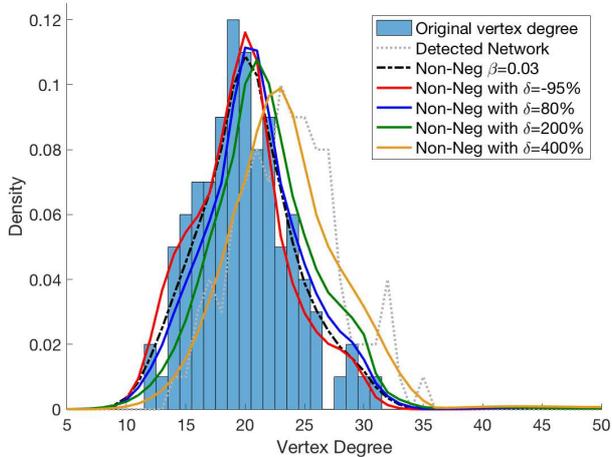}
\caption{Density histogram of the original vertex degrees (blue bars), detected density vertex degree distribution (gray dotted line), and the results when a non-negative constraint is applied to the truncated singular value decomposition using the true $\beta$ (black dashed line), and perturbations from the true $\beta$ (red, blue, yellow, and green solid lines). The original network is an Erd{\H o}s-R{\'e}nyi network with $100$ nodes and probability of connection $0.2$.}
\label{wrongB}
\end{figure}

Figure~\ref{wrongB} shows the reconstruction of an Erd{\H o}s-R{\'e}nyi network with $100$ nodes and probability of a connection of $0.2$ for the true value of $\beta=0.03$, and also for various values of $\beta$ deviating up to $\delta=400\%$ from the true value of $\beta$.
Note that, since $0\leq\beta\leq 1$, the conditions for the perturbations are $-1\leq\delta\leq 1/\beta-1$.
Therefore, the perturbation of $\delta=-95\%$ is almost the most extreme value can be chosen in the negative direction.
The cut-off for the truncated singular value decomposition method is $0.55$ and the probability of \textit{type I error} is $\alpha=0.05$. 
The method performs to be robust to large perturbations of $\beta$, in both negative and positive directions.
Up to $\delta=80\%$ in the positive direction, the bias of the reconstruction is negligible; only if $\delta=400\%$ or more, the reconstruction deviates significantly from the true one, although it still performs better than the na\"ive approach of trusting the identified network structure.

As stated in the concluding section of \citep{GloriaPRE2018}, a limitation of this method is the assumption that the probabilities of \textit{type I} and \textit{type II errors} are known a priori.
It has been shown that wrong estimates of these two errors, within certain bounds, do not cause the reconstruction of be rendered invalid. 
The iterative procedure presented in this paper explores the way to adjust the estimates of these two errors, so to improve the reconstruction of the network of interest.

\section{Iterative procedure}
\label{Procedure}

Each step of the iterative procedure uses the information provided so far to reconstruct the vertex degree distribution of a network, when the biased one is given.

The first step of iteration consists itself of various parts.
After a value of correlation is calculated for each link, the aggregated distribution of all correlation values $(1-p^*) f(N,0,r)+p^* f(N,\rho,r)$ is found, where $p^*$ is the rate of true links over the total.
The aggregated distribution is used to estimate the distributions $f(N,0,r)$ and $f(N,\rho,r)$.
Fix the initial value for $\alpha$ to $0.05$ and find $r_\tau$ using Eq.~(\ref{eqAlpha}).
All the links with corresponding correlation value larger than $r_\tau$ are considered to be present.
The resulting network is the so-called biased network, and therefore the vertex degree distribution 
\begin{equation}
\mathcal{P'}_i=\frac{\text{number of vertices with degree}=i}{\text{total number of vertices}}
\label{empiricalVertexDegree}
\end{equation}
can be calculated empirically by counting the vertices' degrees.
Knowing $N$ and $\rho$, $\beta=\beta(N,\rho,\alpha)$ is found, since $\alpha$ has been fixed, and therefore the matrix $A$ can be evaluated.
All the ingredients needed to solve Eq.~(\ref{pA+p1}) are now available, hence the vertex degree distribution $\mathcal{P}_0$ of the original network is calculated, and this concludes the first iteration step.
Table \ref{scheme} summarises the procedure of the first iteration step.

\begin{table}[!t] 
\centering
\begin{minipage}{0.765\textwidth}
\begin{tabular}{l c r}
\begin{minipage}{0.16\textwidth}
{\color{white} .}
\end{minipage}
\begin{minipage}{0.33\textwidth}
\begin{tcolorbox}[width=0.7\textwidth, height=1cm, valign=center]
Estimate $f(N,\rho,r)$
\end{tcolorbox}
\end{minipage}
\begin{minipage}{0.33\textwidth}
\begin{tikzpicture}[overlay, remember picture, blue!50!black, line width=3pt]
\draw [->] (-1.2,-0.65) to (-0.1,-0.65);
\end{tikzpicture}
\begin{tcolorbox}[width=0.7\textwidth, height=1cm, valign=center]
Fix thresh.

$\alpha=0.05$
\end{tcolorbox}
\end{minipage}\\

\vspace*{1cm}\\

\begin{minipage}{0.33\textwidth}
\begin{tcolorbox}[width=0.7\textwidth, height=1cm, valign=center]
Calculate

$A=A(n,\alpha,\beta)$
\end{tcolorbox}
\end{minipage}
\begin{minipage}{0.33\textwidth}
\begin{tikzpicture}[overlay, remember picture, blue!50!black, line width=3pt]
\draw [<-] (-1.3,-0.65) to (-0.1,-0.65);
\draw [->] (3.6,1.4) to (5.6,0.1);
\draw [->] (3.5,1.4) to (1.5,0.1);
\end{tikzpicture}
\begin{tcolorbox}[width=0.7\textwidth, height=1cm, valign=center]
Find 

$\beta=\beta(N,\rho,\alpha)$
\end{tcolorbox}
\end{minipage}
\begin{minipage}{0.33\textwidth}
\begin{tcolorbox}[width=0.7\textwidth, height=1cm, valign=center]
Calculate distribution $\mathcal{P'}$
\end{tcolorbox}
\end{minipage}\\

\vspace*{1cm} \\

\begin{minipage}{0.33\textwidth}
\begin{tikzpicture}[overlay, remember picture, blue!50!black,shorten >=1pt, line width=3pt]
\draw [->] (1.5,2.0) -- (1.5,1.5) to (5.8,1.5);
\end{tikzpicture}
\end{minipage}
\begin{minipage}{0.33\textwidth}
\begin{tikzpicture}[overlay, remember picture, blue!50!black,shorten >=1pt, line width=3pt]
\draw [->] (1.5,0.9) to (1.5,-0.1);
\end{tikzpicture}
\begin{tcolorbox}[width=0.7\textwidth, height=1cm, valign=center]
Evaluate
$\mathcal{P}_0=A_k^+\ \mathcal{P'}$
\end{tcolorbox}
\end{minipage}
\begin{minipage}{0.33\textwidth}
\begin{tikzpicture}[overlay, remember picture, blue!50!black,shorten >=1pt, line width=3pt]
\draw [->] (1.5,2.0) -- (1.5,1.5) to (-2.8,1.5);
\end{tikzpicture}
\end{minipage}\\
\begin{minipage}{\textwidth}
{\color{white} .}
\end{minipage}
\end{tabular}
\caption{Network Inference: procedure scheme. After a value of correlation is calculated for each link, the aggregated distribution is used to estimate the distributions $f(N,0,r)$ and $f(N,\rho,r)$. Fix the initial value for $\alpha$ to $0.05$, and find $r_\tau$ using Eq.~(\ref{eqAlpha}). All the links with corresponding correlation value larger than $r_\tau$ are considered to be present. The probability density function $\mathcal{P'}$ of the vertex degree distribution of the biased network is calculated. The value for $\beta$ is found using Eq.~(\ref{betaAlphaNrho}), therefore also $A$ is computed. Knowing $\mathcal{P'}$ and $A$, the probability density function $\mathcal{P}_0$ of the original vertex degree distribution is calculated using Eq.~(\ref{pA+p1}).}
\label{scheme}
\end{minipage}
\end{table}

Section \ref{preliminary2} shows that wrong estimates of $\alpha$ and $\beta$, within certain bounds, do not cause the reconstruction $\mathcal{P}=A_k^+\ \mathcal{P'}$ [Eq.~(\ref{pA+p1})] of be rendered invalid.
Note that the robustness of perturbations of $\alpha$ and $\beta$ of the reconstruction method performs differently depending on the network topology of interest.
Therefore, the procedure explained above and summarised in Table~\ref{scheme}, can be enhanced using this idea.
Namely, the density of the vertex degree distribution $\mathcal{P}_0$, that is inferred at the last step of the procedure, is used to perform a robustness analysis varying the value of $\alpha$, and consequently of $\beta$ [Eq.~(\ref{betaAlphaNrho})].
The value of $\alpha$, that gives the most robust result, is used to iterate the procedure.
A general step of iteration is summarised in Table~\ref{scheme2}.

The robustness analysis consists of calculating several different biased networks varying $\alpha$, starting from the vertex degree distribution $\mathcal{P}_0$, i.e., $\mathcal{P'}(\alpha)=A(n,\alpha,\beta)\ \mathcal{P}_0$.
For each $\mathcal{P'}(\alpha)$, a robustness analysis is performed, as explained in Sec.~\ref{preliminary2}.
Namely, call $\mathcal{P}_0^*(\alpha,\delta)$ the density of the reconstructed vertex degree distribution, using as biased vertex degree distribution $\mathcal{P'}(\alpha)$, and a perturbation $\delta$ on $\alpha$.

The value of $\alpha$, that gives the most robust result, is chosen analysing the difference between the distributions $\mathcal{P}_0$ and $\mathcal{P}_0^*(\alpha,\delta)$.
To quantify the bias between these distributions, the Kolmogorov-Smirnov distance is considered.
The Kolmogorov-Smirnov test is a test based on the closeness of two distributions, which uses the Kolmogorov-Smirnov statistic to perform a test of hypothesis.
The Kolmogorov-Smirnov distance is the largest difference between the two cumulative distributions \citep{quinn2002}, i.e.,

\begin{equation}
\mathcal{D}_{\alpha,\delta}=\mathcal{D}(\mathcal{F}_0^*(\alpha,\delta),\mathcal{F}_0)=\max |\mathcal{F}_0^*(\alpha,\delta)-\mathcal{F}_0|,
\label{distKolmo}
\end{equation}
is the distance between $\mathcal{P}_0$ and $\mathcal{P}_0^*(\alpha,\delta)$, where $\mathcal{F}_0$ and $\mathcal{F}_0^*(\alpha,\delta)$ are their respective cumulative distributions.
In mathematics, defined on a vector space, this distance is called Tchebychev distance, or maximum metric.

Once the distance $\mathcal{D}_{\alpha,\delta}$ between the two distributions is calculated, the value for $\alpha$ which gives the shortest distance varying $\delta$ is chosen.
This approach follows the minimum distance estimation method developed in \citep{Drossos1980,Parr1980,Wolfowitz1952,Matusita1953,wolfowitz1957}.

\begin{table}[!t]
\centering
\begin{minipage}{0.765\textwidth}
\begin{tabular}{l c r}
\begin{minipage}{0.16\textwidth}
{\color{white} .}
\end{minipage}
\begin{minipage}{0.33\textwidth}
\begin{tcolorbox}[width=0.7\textwidth, height=1cm, valign=center]
Estimate corr. coeff. $\rho$
\end{tcolorbox}
\end{minipage}
\begin{minipage}{0.33\textwidth}
\begin{tikzpicture}[overlay, remember picture, blue!50!black,shorten >=1pt, line width=3pt]
\draw [->] (-1.2,-0.65) to (-0.1,-0.65);
\end{tikzpicture}
\begin{tcolorbox}[width=0.7\textwidth, height=1cm, valign=center]
Fix thresh.

{\color{red!80!black} $\alpha=\alpha(\mathcal{P}_0)$}
\end{tcolorbox}
\end{minipage}
\begin{minipage}{0.1\textwidth}
\begin{tikzpicture}[overlay, remember picture, red!80!black,shorten >=1pt, line width=3pt]
\draw [->] (-3.3,-5.4) -- node[above] {Robustness Analysis} (1.2,-5.4) -- (1.2,0) to (-1.3,0);
\end{tikzpicture}
\end{minipage}\\

\vspace*{1cm}\\

\begin{minipage}{0.33\textwidth}
\begin{tcolorbox}[width=0.7\textwidth, height=1cm, valign=center]
Calculate

$A=A(n,\alpha,\beta)$
\end{tcolorbox}
\end{minipage}
\begin{minipage}{0.33\textwidth}
\begin{tikzpicture}[overlay, remember picture, blue!50!black,shorten >=1pt, line width=3pt]
\draw [<-] (-1.3,-0.65) to (-0.1,-0.65);
\draw [->] (3.6,1.4) to (5.6,0.1);
\draw [->] (3.5,1.4) to (1.5,0.1);
\end{tikzpicture}
\begin{tcolorbox}[width=0.7\textwidth, height=1cm, valign=center]
Find

$\beta=\beta(N,\rho,\alpha)$
\end{tcolorbox}
\end{minipage}
\begin{minipage}{0.33\textwidth}
\begin{tcolorbox}[width=0.7\textwidth, height=1cm, valign=center]
Calculate distribution $\mathcal{P'}$
\end{tcolorbox}
\end{minipage}\\

\vspace*{1cm} \\

\begin{minipage}{0.33\textwidth}
\begin{tikzpicture}[overlay, remember picture, blue!50!black,shorten >=1pt, line width=3pt]
\draw [->] (1.5,2.0) -- (1.5,1.5) to (5.8,1.5);
\end{tikzpicture}
\end{minipage}
\begin{minipage}{0.33\textwidth}
\begin{tikzpicture}[overlay, remember picture, blue!50!black,shorten >=1pt, line width=3pt]
\draw [->] (1.5,0.9) to (1.5,-0.1);
\end{tikzpicture}
\begin{tcolorbox}[width=0.7\textwidth, height=1cm, valign=center]
Evaluate
{\color{red!80!black} $\mathcal{P}_0=A_k^+\ \mathcal{P'}$}
\end{tcolorbox}
\end{minipage}
\begin{minipage}{0.33\textwidth}
\begin{tikzpicture}[overlay, remember picture, blue!50!black,shorten >=1pt, line width=3pt]
\draw [->] (1.5,2.0) -- (1.5,1.5) to (-2.8,1.5);
\end{tikzpicture}
\end{minipage}\\
\begin{minipage}{\textwidth}
{\color{white} .}
\end{minipage}
\end{tabular}
\caption{Network Inference: procedure scheme with iteration. After a value of correlation is calculated for each link, the aggregated distribution is used to estimate the distributions $f(N,0,r)$ and $f(N,\rho,r)$. Fix the value for $\alpha$ according to the result of the previous iteration step, and find $r_\tau$ using Eq.~(\ref{eqAlpha}). All the links with corresponding correlation value larger than $r_\tau$ are considered to be present. The probability density function $\mathcal{P'}$ of the vertex degree distribution of the biased network is calculated. The value for $\beta$ is found using Eq.~(\ref{betaAlphaNrho}), therefore also $A$ is computed. Knowing $\mathcal{P'}$ and $A$, the probability density function $\mathcal{P}_0$ of the original vertex degree distribution is calculated using Eq.~(\ref{pA+p1}). Perform a robustness analysis to iterate the procedure.}
\label{scheme2}
\end{minipage}
\end{table}

\section{Simulation study, results and discussion}
\label{ResultsProcedure}

In this section a simulation study is presented to show the applicability of the method described in the previous section.
Consider an Erd{\H o}s-R{\'e}nyi network with $n=100$ vertices and probability of connection $p=0.2$.
For the sake of simplicity, assume this is a dynamical system, and the dynamics of the nodes are accessible.
Performing a correlation analysis for every pair of vertices, values for the correlation are found.
As explained in Sec.~\ref{preliminary1}, it is reasonable to assume that correlation values corresponding to connected nodes follow the density distribution $f(N,\rho,r)$ [Eq.~(\ref{correlation})].
Therefore, instead of simulating the dynamics of the nodes, a value for the correlation is assigned at random with density distribution $f(N,\rho,r)$ to each link, with $\rho=0.25$ and the number of data points $N=100$.
For each absent link the value for the correlation is chosen at random with distribution $f(N,0,r)$.
The aggregated distribution of all correlation values is used to estimate the true correlation coefficient $\rho$.

\begin{figure}[!t]
\centering
\begin{minipage}{0.46\textwidth}
\flushleft
\includegraphics[scale=0.4]{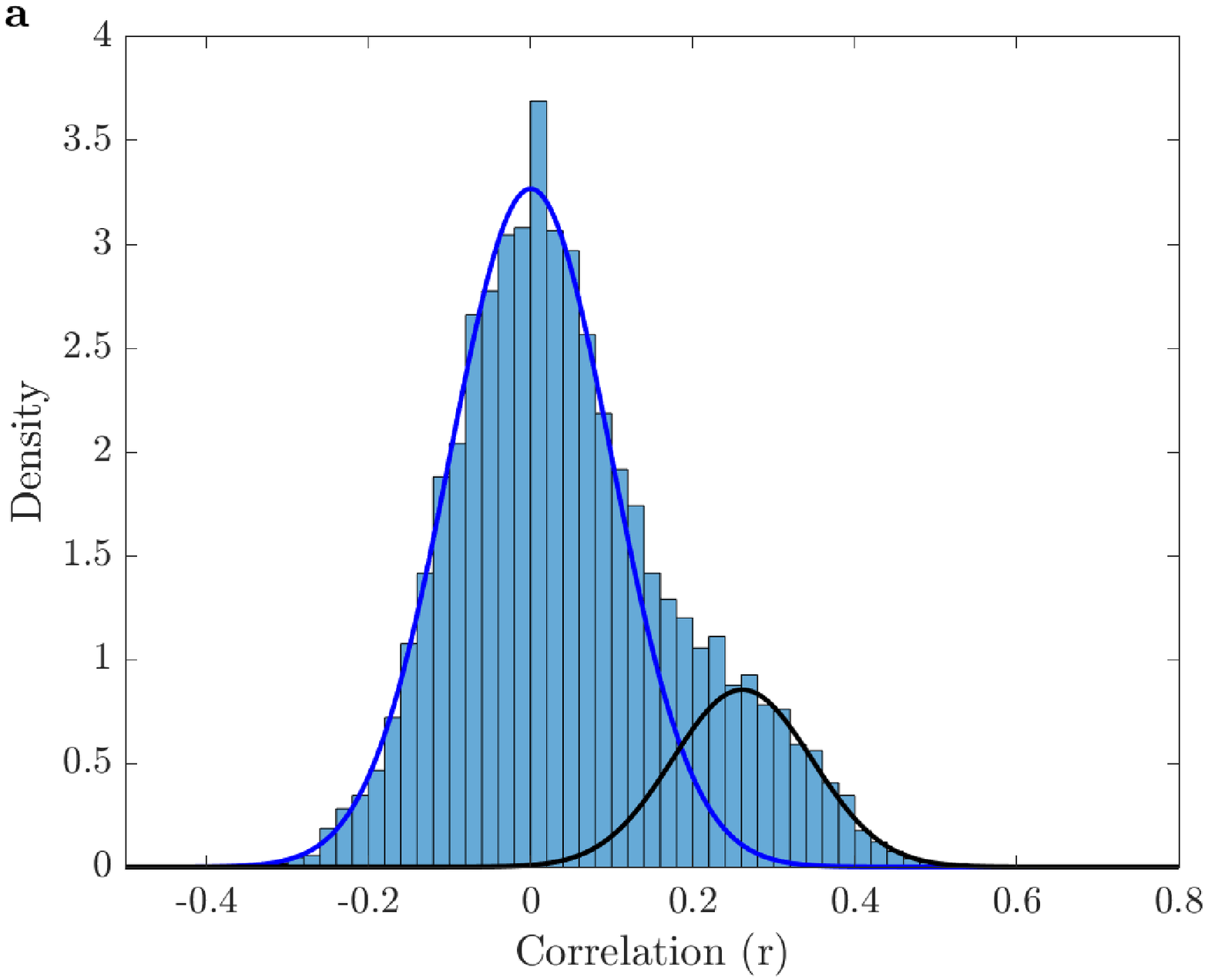}
\end{minipage}
\begin{minipage}{0.46\textwidth}
\flushright
\includegraphics[scale=0.4]{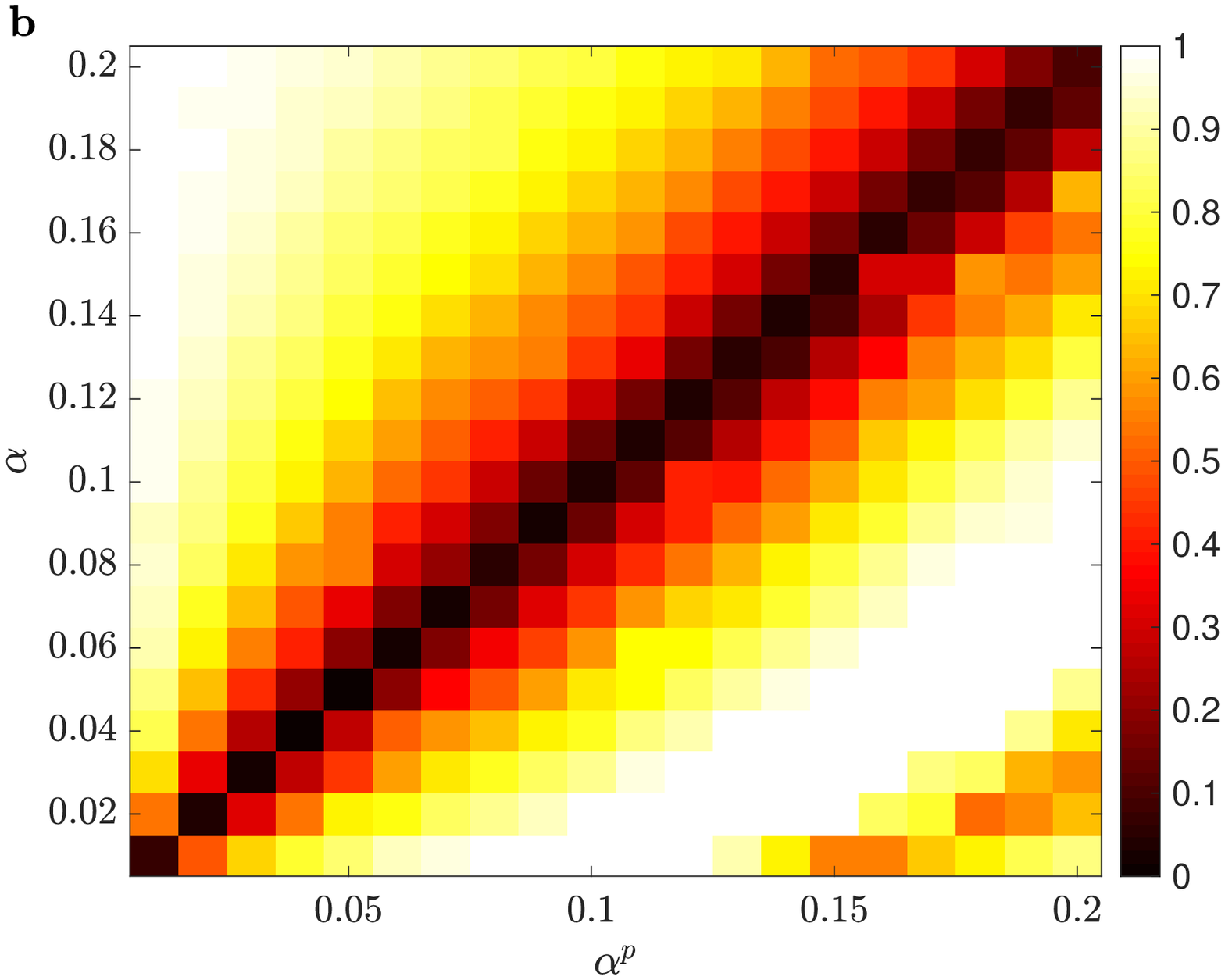}
\end{minipage}
\begin{minipage}{0.02\textwidth}
\centering
\rotatebox{90}{$\mathcal{D}_{\alpha,\alpha^p}$}
\end{minipage} \qquad
\begin{minipage}{0.46\textwidth}
\flushleft
\includegraphics[scale=0.4]{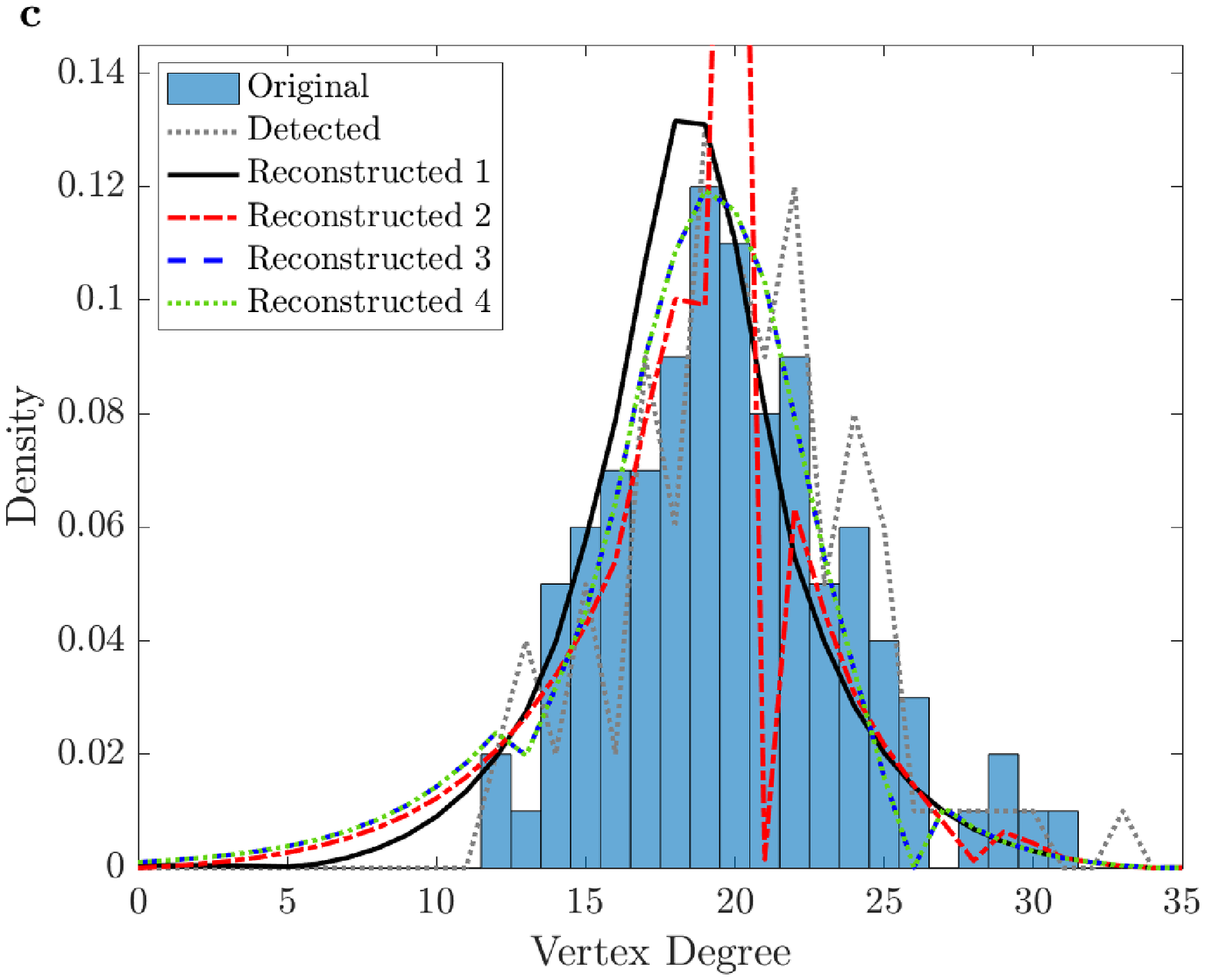}
\end{minipage}
\begin{minipage}{0.46\textwidth}
\flushleft
\includegraphics[scale=0.4]{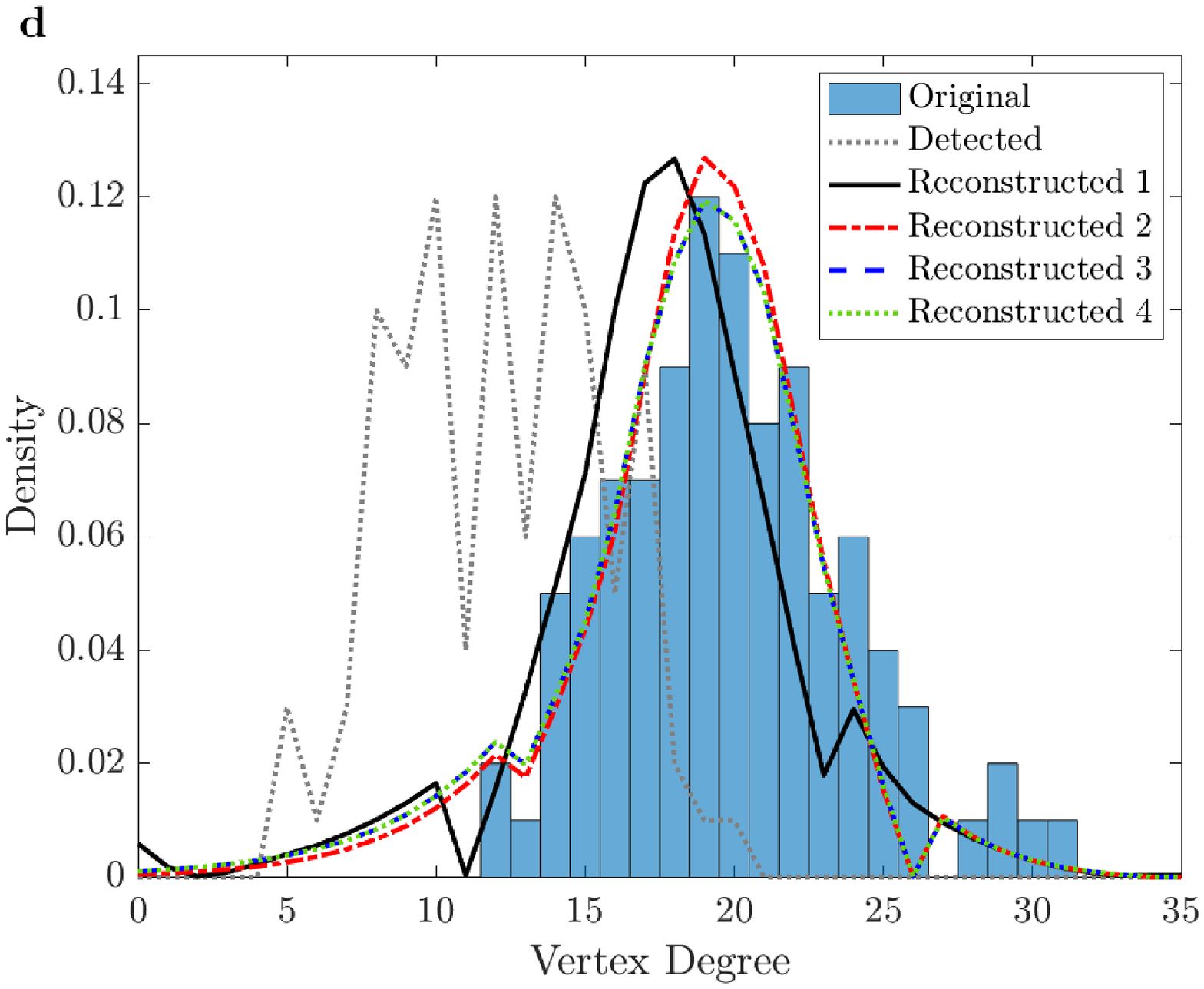}
\end{minipage}
\caption{Reconstruction of an Erd{\H o}s-R{\'e}nyi network with $n=100$ vertices and probability of connection $0.2$. a) Histogram of all correlation values, and the estimated distributions, using the expectation maximization algorithm for Gaussian mixture models. b) Kolmogorov-Smirnov distance $\mathcal{D}_{\alpha,\alpha^p}$ obtained from $\mathcal{D}_{\alpha,\delta}$, where $\alpha^p=\alpha+\delta\alpha$ (x-axis).  c) Results of four iteration steps, starting with $\alpha=0.05$. d) Results of four iteration steps, starting with $\alpha=0.01$.}
\label{figIteration}
\end{figure}

In real world applications, the expectation maximization algorithm for Gaussian mixture models is often used to estimate two distributions given the overall observed data.
This algorithm is an iterative procedure to find maximum likelihood estimates of the parameters in the model.
The expectation maximization algorithm for a Gaussian mixture model is used in this paper to estimate the distributions for true and absent links of the correlation coefficients.
The choice of fitting normal distributions, instead of the exact distribution equation [Eq.~(\ref{correlation})], is motivated by the fact that we want to keep our approach as general as possible.
Figure~\ref{figIteration}a shows the histogram of all correlation values, and the estimated distributions, using the expectation maximization algorithm for Gaussian mixture models.

Once the distributions for true and absent links are estimated, and $\alpha$ is set to $0.05$, all the correlation values above the associated threshold $r_\tau$ are selected; the corresponding links are considered to be present, and the biased, or detected, network is obtained.
The procedure explained in Sec.~\ref{Procedure} is now applied to find the vertex degree distribution of the original network.

The result of the first iteration $\mathcal{P}_0=A_k^+\ \mathcal{P'}$ is used to perform a robustness analysis and the densities $\mathcal{P}_0^*(\alpha,\delta)$ of the vertex degree distributions are calculated.
Evaluate the Kolmogorov-Smirnov distance $\mathcal{D}_{\alpha,\delta}$ using Eq.~(\ref{distKolmo}).
For the sake of simplicity in the representation of the results, the value of the perturbed $\alpha$, i.e., $\alpha^p=\alpha+\delta\alpha$ is used instead of $\delta$, and the distance $\mathcal{D}_{\alpha,\alpha^p}$ is plotted, as in Fig.~\ref{figIteration}b.

Once $\mathcal{D}_{\alpha,\delta}$ is evaluated, the value for $\alpha$ which gives the shortest distance varying $\delta$ is chosen.
Namely, the aim is to minimise the bias of the reconstruction in a neighbourhood of $\alpha$. 
For the same values of the perturbation $\delta$, the distance $\mathcal{D}_{\alpha,\delta}$ gives different results, depending on $\alpha$.
This means that there exist some values for $\alpha$ which give smaller values for the overall distance $\mathcal{D}_{\alpha,\delta}$ in a neighbourhood of $\alpha$.
Once the radius of the neighbourhood is chosen, e.g. $-0.2<\delta<0.2$ , the overall bias of the neighbourhood is calculated as $\sum_{-0.2<\delta<0.2}(1-\delta)\mathcal{D}_{\alpha,\delta}$, the value for $\alpha$ which gives the smallest result is chosen, and the procedure is iterated.

Figure~\ref{figIteration}c shows the results of four iteration steps.
As shown in~\ref{preliminary1}, small perturbations of $\alpha$ and $\beta$ do not cause the failure of the reconstruction method; in this example, this is reflected by the fact that the result of the first iteration gives already a correct reconstruction.
The results for the last two iteration steps demonstrate that the process converges, and once the correct reconstruction is achieved, the following results do not deviate from it.

Figure~\ref{figIteration}d shows the results of four iteration steps, when the initial value of $\alpha$ is set to 0.01 (instead of 0.05).
In this case, the detected vertex degree distribution deviates considerably from the original one, implying that the na\"ive approach would lead to a wrong estimate of the vertex degree distribution.
The result of the first iteration improves the result, reducing the bias and getting the correct shape of the distribution.
At the other iteration steps, the bias is completely corrected.
Note that the last two iteration steps give the same reconstructed distribution, this means that the procedure converged to a solution and the following results do not deviate from it.

\begin{figure}[!t]
\centering
\begin{minipage}{0.46\textwidth}
\flushleft
\includegraphics[scale=0.4]{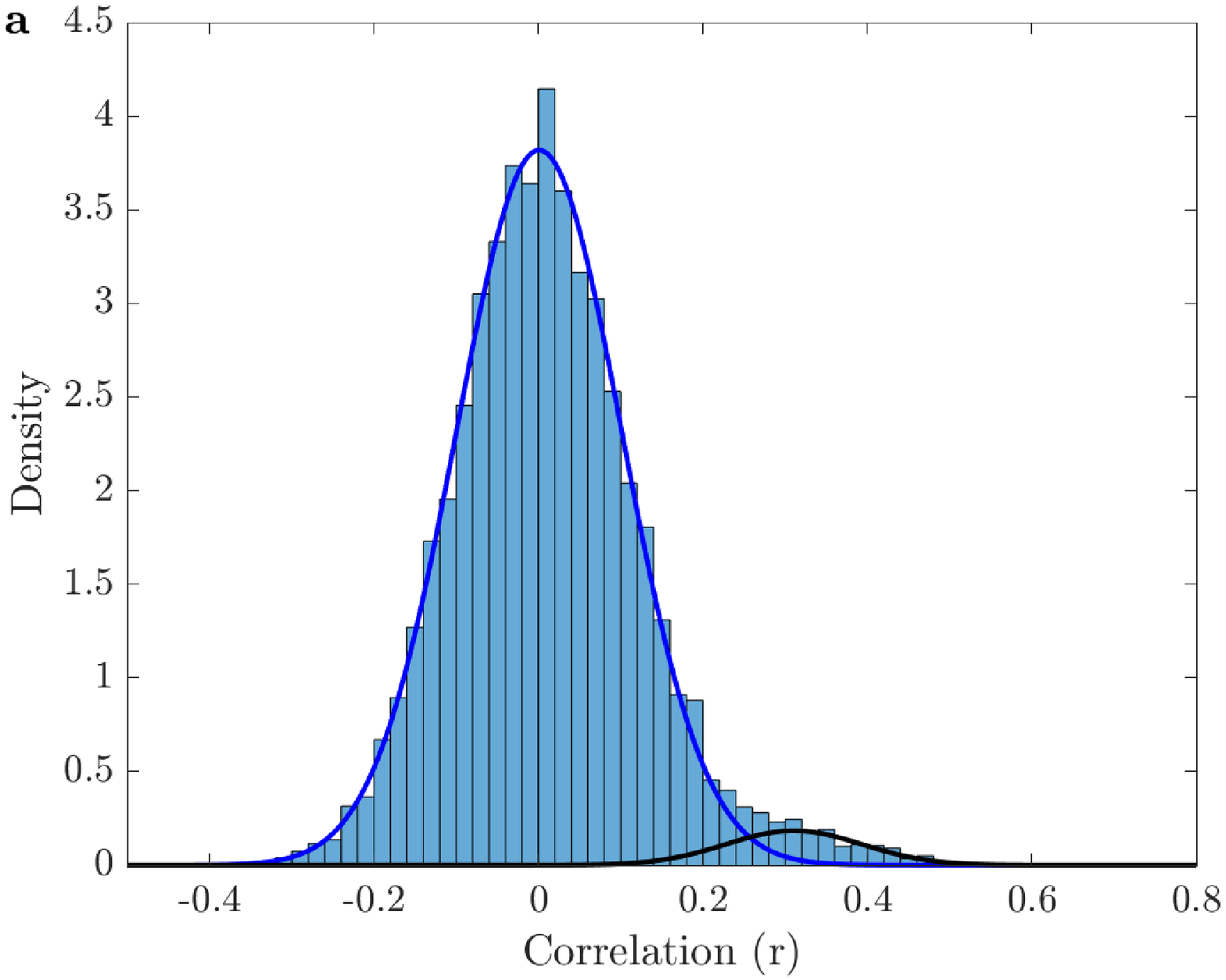}
\end{minipage}
\begin{minipage}{0.46\textwidth}
\flushright
\includegraphics[scale=0.4]{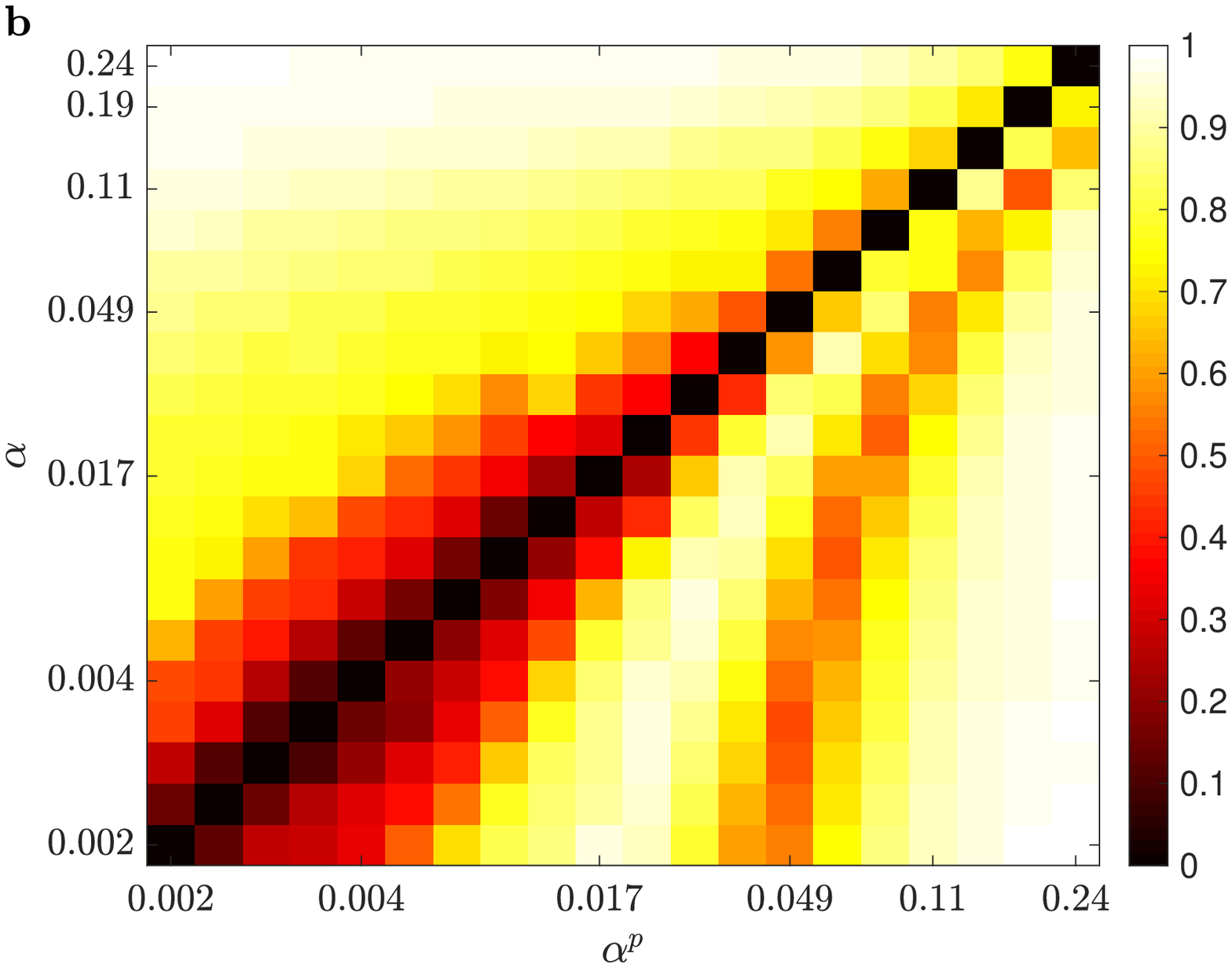}
\end{minipage}
\begin{minipage}{0.02\textwidth}
\centering
\rotatebox{90}{$\mathcal{D}_{\alpha,\alpha^p}$}
\end{minipage} \qquad
\begin{minipage}{0.46\textwidth}
\flushleft
\includegraphics[scale=0.4]{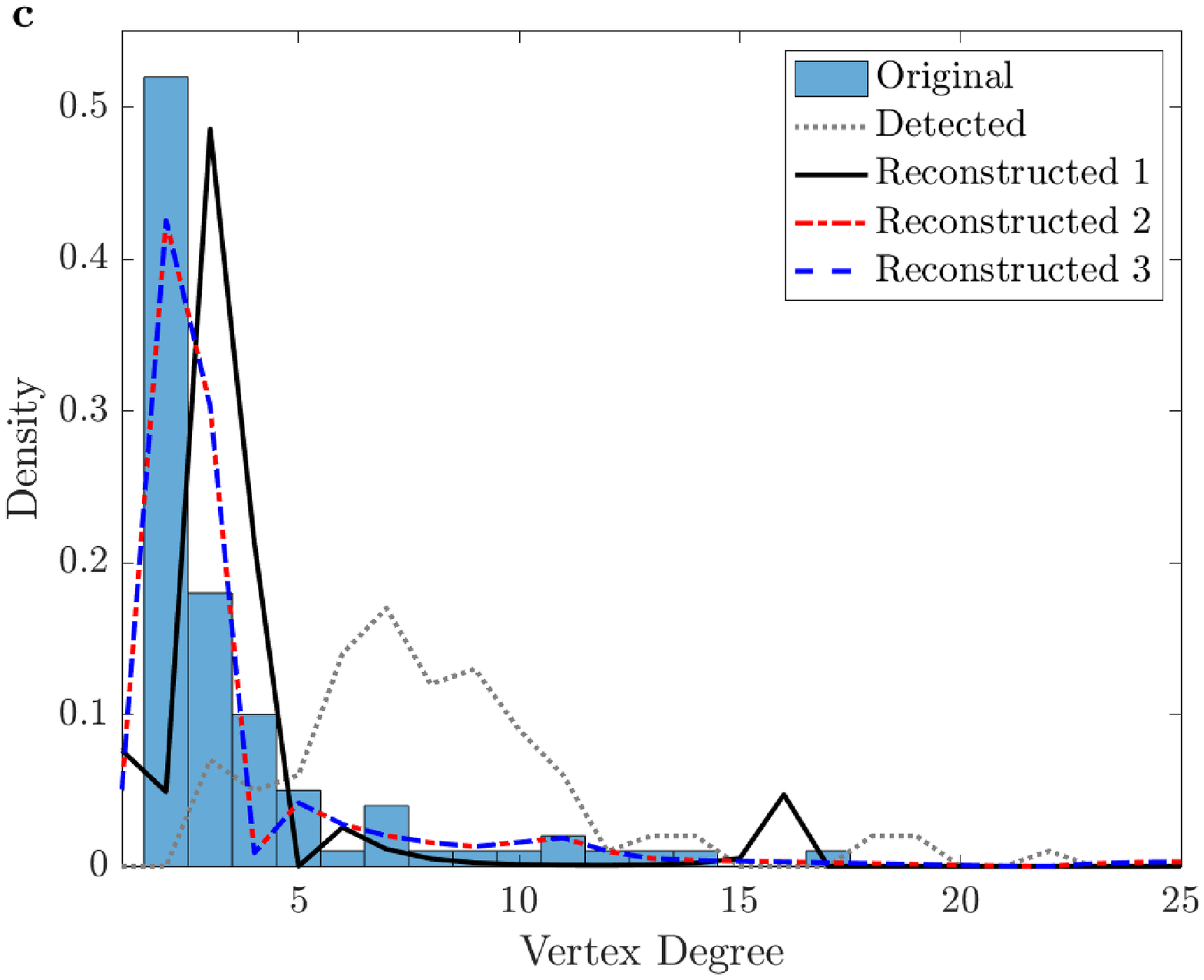}
\end{minipage}
\begin{minipage}{0.46\textwidth}
\flushleft
\includegraphics[scale=0.4]{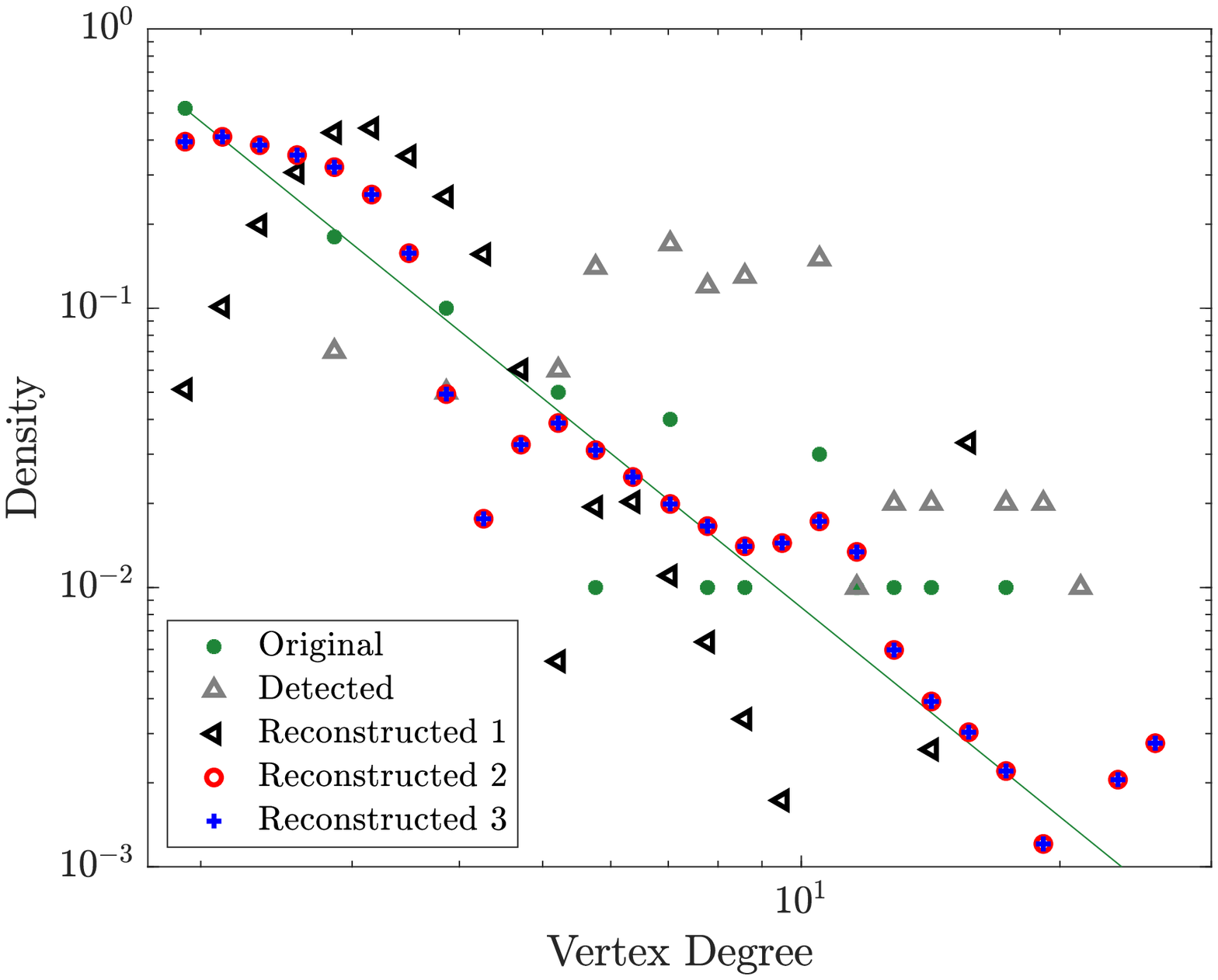}
\end{minipage}
\caption{Reconstruction of a scale-free network with $n=100$ vertices. a) Histogram of all correlation values, and the estimated distributions, using the expectation maximization algorithm for Gaussian mixture models. b) Kolmogorov-Smirnov distance $\mathcal{D}_{\alpha,\alpha^p}$ obtained from $\mathcal{D}_{\alpha,\delta}$, where $\alpha^p=\alpha+\delta\alpha$ (x-axis). c) Results of three iteration steps with initial $\alpha=0.05$. d) Results of three iteration steps presented in a log-log plot.}
\label{figIterationScale}
\end{figure}

The method described in the previous section is now applied to a Scale-Free network of 100 vertices.
Values for the correlation are assigned at random with density distribution $f(N,\rho,r)$ to each link, with $\rho=0.3$ and the number of data points $N=100$.
Figure~\ref{figIterationScale}a shows the histogram of all correlation values, and the estimated distributions, using the expectation maximization algorithm for Gaussian mixture models.
Figure~\ref{figIterationScale}b depicts the Kolmogorov-Smirnov distance $\mathcal{D}_{\alpha,\alpha^p}$ obtained after the first iteration step.
Interestingly, the values for $\alpha$ that give the most robust results are obtained for $\alpha$ as small as possible; intuitively, this means that high certainty about the presence of links is needed to recover the scale-freeness property of a network.
A non-linear x- and y-scale for $\alpha$ is here used to emphasize the role of small values for $\alpha$ on the distance $\mathcal{D}_{\alpha,\alpha^p}$.
Note that this is in agreement with the results shown in Fig.~5 in \citep{GloriaJNM2018}; in that case, the relation between $\alpha$ and $\beta$ is different from the case presented in this section, nevertheless, the general concept of keeping $\alpha$ as small as possible to correctly recover the scale-freeness property, remains valid.
Figure~\ref{figIterationScale}c shows the results of three iteration steps, when the initial value of $\alpha$ is set to 0.05.
Also in this case, the last two iteration steps give the same reconstructed distribution, meaning that the procedure converged to a solution and the following results do not deviate from it.
Figure~\ref{figIterationScale}d presents the plot in Fig.~\ref{figIterationScale}c in log-log scale to demonstrate that the scale-freeness property of the network has been correctly recovered.

Additionally, simulations with both larger and smaller number of nodes have been performed.
Results are qualitatively the same as the ones presented in this section.
Likewise, for different values of $\rho$ the method performs similarly.
However, for very small values of $\rho$, the aggregated distribution of all correlation values does not carry any relevant information about the distributions of true and absent links.
Namely, the distribution of the true links is shifted towards zero, and therefore it is superposed to the other one. 
Consequently, the expectation maximization algorithm for Gaussian mixture models is not accurate in distinguishing the two distributions for the correlation coefficients.
Hence, the correct reconstruction of the network is not guaranteed.
In this case, i.e., when the aggregated distribution of all correlation values does not carry any relevant information, different measures for the strength of connection should be used instead of correlation.

\section{Conclusion}
\label{ConclusionProcedure}

The analysis presented in \citep{GloriaPRE2018} is advanced in this manuscript, and an iterative procedure to reconstruct the vertex degree distribution is suggested.
This procedure should be used when reconstructing a network from data to guarantee that the vertex degree distribution is correctly recovered.
In particular, when the estimates of $\alpha$ and $\beta$ might not be accurate and large errors on these estimates might occur, the method described in this manuscript ensures a robust and better reconstruction than other methods described previously.

This iteration procedure allows to gain, at each step, a better insight on the network topology.
Consequently, the value for $\alpha$ can be tuned to obtain a more robust reconstruction, thereby achieving a better result.
It is shown, using some examples, that this method converges to the correct reconstruction of the vertex degree distribution.
Furthermore, only few iteration steps are needed to achieve this goal.

The method presented in this manuscript is a general procedure that can be applied to any kind of network.
Moreover, this iterative procedure does not depend on the dynamic of the nodes, nor on the dimension of the network.
Throughout the paper correlation has been used as a measure of connectivity strength for each link; nevertheless, the method works regardless the kind of measure chosen to establish the strength of connection.
This implies that different types of measures can be used instead of correlation, broadening the applicability of the procedure described in this work.

Future investigations should apply this iterative procedure to various network topologies, to demonstrate that these results are not only case specific.
Moreover, future studies should analyse the robustness of this procedure and find the size of perturbation needed to make the reconstruction fail even after various iteration steps.
Furthermore, with the aim to make this procedure even more general, the case in which the inference of the distributions for the correlation coefficients is not accurate should be carefully studied.

Note that the method presented in this manuscript aims to reconstruct the vertex degree distribution.
A generalization for other characteristics is likely more complicated, and therefore requires a more in-depth analysis.
Nonetheless, this further investigation could lead to reconstructing the network itself, i.e., link by link estimation.
Namely, once the functional relationship between the reconstructed and true underlying network is found in terms of vertex degree distribution and eventually also, e.g., the shortest path length and the clustering coefficient, the true underlying network should be inferred combining all these results.
A step further consists of creating a whole new procedure for network inference.
Firstly, the vertex degree distribution and eventually other characteristics of the original network are estimated.
Secondly, the links in the network which have zero probability of being false positive or false negative are identified. 
Such links are fixed, and the method aims to reconstruct the remaining links matching the characteristics found at the previous step.

\section*{Acknowledgements}

This project has received funding from the European Union's Horizon 2020 research and innovation programme under the Marie Sklodowska-Curie grant agreement No 642563.

The authors declare that they have no known competing financial interests or personal relationships that could have appeared to influence the work reported in this paper.

\section*{References}
\bibliographystyle{elsarticle-num}
\bibliography{biblioThesis.bib}

\begin{thebibliography}{10}
\expandafter\ifx\csname url\endcsname\relax
  \def\url#1{\texttt{#1}}\fi
\expandafter\ifx\csname urlprefix\endcsname\relax\def\urlprefix{URL }\fi
\expandafter\ifx\csname href\endcsname\relax
  \def\href#1#2{#2} \def\path#1{#1}\fi

\bibitem{Barrat2008}
B.~A., B.~M., V.~A., Dynamical Processes on Complex Networks, Cambridge
  University Press, Cambridge, 2008.

\bibitem{Boccaletti2006}
S.~Boccaletti, V.~Latora, Y.~Moreno, M.~Chavez, D.-U. Hwang, Complex networks:
  Structure and dynamics, Phys. Rep. 424.
\newblock \href
  {http://dx.doi.org/https://doi.org/10.1016/j.physrep.2005.10.009}
  {\path{doi:https://doi.org/10.1016/j.physrep.2005.10.009}}.

\bibitem{Cohen2010}
R.~Cohen, S.~Havlin, Complex Networks: Structure, Robustness and Function,
  Cambridge University Press, Cambridge, 2010.

\bibitem{Fagiolo2007}
G.~Fagiolo, Clustering in complex directed networks, Phys. Rev. E 76.
\newblock \href {http://dx.doi.org/10.1103/PhysRevE.76.026107}
  {\path{doi:10.1103/PhysRevE.76.026107}}.

\bibitem{Barabasi2002}
R.~Albert, A.~L. Barab\'asi, Statistical mechanics of complex networks, Rev.
  Mod. Phys. 74.
\newblock \href {http://dx.doi.org/10.1103/RevModPhys.74.47}
  {\path{doi:10.1103/RevModPhys.74.47}}.

\bibitem{Banavar}
J.~R. Banavar, F.~Colaiori, A.~Flammini, A.~Maritan, A.~Rinaldo, Topology of
  the fittest transportation network, Phys. Rev. Lett. 84.
\newblock \href {http://dx.doi.org/10.1103/PhysRevLett.84.4745}
  {\path{doi:10.1103/PhysRevLett.84.4745}}.

\bibitem{Bar2003}
Y.~Bar-Yam, Dynamics of complex systems, Boulder, CO : Westview Press, 2003.

\bibitem{Dorogovtsev2003}
S.~N. Dorogovtsev, J.~F.~F. Mendes, {Evolution of Networks: From Biological
  Nets to the Internet and WWW}, Oxford University Press, Oxford, 2003.
\newblock \href {http://dx.doi.org/10.1093/acprof:oso/9780198515906.001.0001}
  {\path{doi:10.1093/acprof:oso/9780198515906.001.0001}}.

\bibitem{Egu}
V.~M. Egu\'{\i}luz, D.~R. Chialvo, G.~A. Cecchi, M.~Baliki, A.~V. Apkarian,
  Scale-free brain functional networks, Phys. Rev. Lett. 94.
\newblock \href {http://dx.doi.org/10.1103/PhysRevLett.94.018102}
  {\path{doi:10.1103/PhysRevLett.94.018102}}.

\bibitem{Kurant}
M.~Kurant, P.~Thiran, Extraction and analysis of traffic and topologies of
  transportation networks, Phys. Rev. E 74.
\newblock \href {http://dx.doi.org/10.1103/PhysRevE.74.036114}
  {\path{doi:10.1103/PhysRevE.74.036114}}.

\bibitem{Newman2003}
M.~E.~J. Newman, The structure and function of complex networks, SIAM Rev. 45.
\newblock \href
  {http://arxiv.org/abs/http://dx.doi.org/10.1137/S003614450342480}
  {\path{arXiv:http://dx.doi.org/10.1137/S003614450342480}}, \href
  {http://dx.doi.org/10.1137/S003614450342480}
  {\path{doi:10.1137/S003614450342480}}.

\bibitem{Odor}
G.~\'Odor, B.~Hartmann, Heterogeneity effects in power grid network models,
  Phys. Rev. E 98.
\newblock \href {http://dx.doi.org/10.1103/PhysRevE.98.022305}
  {\path{doi:10.1103/PhysRevE.98.022305}}.

\bibitem{Valencia}
M.~Valencia, J.~Martinerie, S.~Dupont, M.~Chavez, Dynamic small-world behavior
  in functional brain networks unveiled by an event-related networks approach,
  Phys. Rev. E 77.
\newblock \href {http://dx.doi.org/10.1103/PhysRevE.77.050905}
  {\path{doi:10.1103/PhysRevE.77.050905}}.

\bibitem{Xue}
Y.~Xue, J.~Wang, L.~Li, D.~He, B.~Hu, Optimizing transport efficiency on
  scale-free networks through assortative or dissortative topology, Phys. Rev.
  E 81.
\newblock \href {http://dx.doi.org/10.1103/PhysRevE.81.037101}
  {\path{doi:10.1103/PhysRevE.81.037101}}.

\bibitem{Yeung}
C.~H. Yeung, K.~Y.~M. Wong, Phase transitions in transportation networks with
  nonlinearities, Phys. Rev. E 80.
\newblock \href {http://dx.doi.org/10.1103/PhysRevE.80.021102}
  {\path{doi:10.1103/PhysRevE.80.021102}}.

\bibitem{Rok2017}
R.~Cestnik, M.~Rosenblum, Reconstructing networks of pulse-coupled oscillators
  from spike trains, Phys. Rev. E 96.
\newblock \href {http://dx.doi.org/10.1103/PhysRevE.96.012209}
  {\path{doi:10.1103/PhysRevE.96.012209}}.

\bibitem{Arkady}
B.~Kralemann, A.~Pikovsky, M.~Rosenblum, Reconstructing effective phase
  connectivity of oscillator networks from observations, New J. Phys. 16.

\bibitem{Li2014}
S.~Li, F.~Li, W.~Liu, M.~Zhan, Network reconstruction by linear dynamics,
  Physica A 404.

\bibitem{Arkady2016}
A.~Pikovsky, Reconstruction of a neural network from a time series of firing
  rates, Phys. Rev. E 93.
\newblock \href {http://dx.doi.org/10.1103/PhysRevE.93.062313}
  {\path{doi:10.1103/PhysRevE.93.062313}}.

\bibitem{Olbrich2010}
E.~Olbrich, T.~Kahle, N.~Bertschinger, N.~Ay, J.~Jost, Quantifying structure in
  networks, Eur. Phys. J. B 77.
\newblock \href {http://dx.doi.org/10.1140/epjb/e2010-00209-0}
  {\path{doi:10.1140/epjb/e2010-00209-0}}.

\bibitem{NewmanBook}
M.~E.~J. Newman, Networks: An Introduction, Oxford University Press, Inc., New
  York, NY, USA, 2010.

\bibitem{Asllani2018}
M.~Asllani, T.~Carletti, F.~Di~Patti, D.~Fanelli, F.~Piazza, Hopping in the
  crowd to unveil network topology, Phys. Rev. Lett. 120 (2018) 158301.
\newblock \href {http://dx.doi.org/10.1103/PhysRevLett.120.158301}
  {\path{doi:10.1103/PhysRevLett.120.158301}}.

\bibitem{Burioni2014}
R.~Burioni, M.~Casartelli, M.~di~Volo, R.~Livi, A.~Vezzani, Average synaptic
  activity and neural networks topology: a global inverse problem, Sci. Rep.
  4~(1) (2014) 4336.
\newblock \href {http://dx.doi.org/10.1038/srep04336}
  {\path{doi:10.1038/srep04336}}.

\bibitem{Shandilya2011}
S.~G. Shandilya, M.~Timme, Inferring network topology from complex dynamics,
  New J. Phys. 13~(1) (2011) 013004.
\newblock \href {http://dx.doi.org/10.1088/1367-2630/13/1/013004}
  {\path{doi:10.1088/1367-2630/13/1/013004}}.

\bibitem{Leguia2019}
M.~G. Leguia, Z.~Levnajić, L.~Todorovski, B.~Ženko, Reconstructing dynamical
  networks via feature ranking, Chaos 29~(9) (2019) 093107.
\newblock \href {http://dx.doi.org/10.1063/1.5092170}
  {\path{doi:10.1063/1.5092170}}.

\bibitem{Banerjee2019}
A.~Banerjee, J.~Pathak, R.~Roy, J.~G. Restrepo, E.~Ott, Using machine learning
  to assess short term causal dependence and infer network links, Chaos 29~(12)
  (2019) 121104.
\newblock \href {http://dx.doi.org/10.1063/1.5134845}
  {\path{doi:10.1063/1.5134845}}.

\bibitem{Panaggio2019}
M.~J. Panaggio, M.-V. Ciocanel, L.~Lazarus, C.~M. Topaz, B.~Xu, Model
  reconstruction from temporal data for coupled oscillator networks, Chaos
  29~(10) (2019) 103116.
\newblock \href {http://dx.doi.org/10.1063/1.5120784}
  {\path{doi:10.1063/1.5120784}}.

\bibitem{Chavez2010}
M.~Chavez, M.~Valencia, V.~Latora, J.~Martinerie, Complex networks: new trends
  for the analysis of brain connectivity, Int. J. Bifurcat. Chaos 20.
\newblock \href {http://dx.doi.org/10.1142/S0218127410026757}
  {\path{doi:10.1142/S0218127410026757}}.

\bibitem{Fallani2014}
F.~De~Vico~Fallani, J.~Richiardi, M.~Chavez, S.~Achard, Graph analysis of
  functional brain networks: practical issues in translational neuroscience,
  Philos. T. R. Soc. B 369.
\newblock \href {http://dx.doi.org/10.1098/rstb.2013.0521}
  {\path{doi:10.1098/rstb.2013.0521}}.

\bibitem{devore2011}
J.~L. Devore, Probability and Statistics for Engineering and the Sciences, 8th
  Edition, Brooks/Cole, 2011, iSBN-13: 978-0-538-73352-6.

\bibitem{Honey2007}
C.~J. Honey, R.~K{\"o}tter, M.~Breakspear, O.~Sporns, Network structure of
  cerebral cortex shapes functional connectivity on multiple time scales, Proc.
  Natl. Acad. Sci. 104.
\newblock \href {http://dx.doi.org/10.1073/pnas.0701519104}
  {\path{doi:10.1073/pnas.0701519104}}.

\bibitem{jalili2011}
M.~Jalili, M.~G. Knyazeva, Constructing brain functional networks from eeg:
  partial and unpartial correlations., J. Integr. Neurosci. 10.

\bibitem{quinn2002}
G.~P. Quinn, M.~J. Keough, Experimental Design and Data Analysis for
  Biologists, Cambridge University Press, 2002.

\bibitem{Schinkel2011}
S.~Schinkel, G.~Zamora-L\'opez, O.~Dimigen, W.~Sommer, J.~Kurths, Functional
  network analysis reveals differences in the semantic priming task, J.
  Neurosci. Meth. 197.
\newblock \href
  {http://dx.doi.org/https://doi.org/10.1016/j.jneumeth.2011.02.018}
  {\path{doi:https://doi.org/10.1016/j.jneumeth.2011.02.018}}.

\bibitem{GloriaJNM2018}
G.~Cecchini, M.~Thiel, B.~Schelter, L.~Sommerlade, Improving network inference:
  The impact of false positive and false negative conclusions about the
  presence or absence of links, J. Neurosci. Meth. 307.
\newblock \href
  {http://dx.doi.org/https://doi.org/10.1016/j.jneumeth.2018.06.011}
  {\path{doi:https://doi.org/10.1016/j.jneumeth.2018.06.011}}.

\bibitem{GloriaPRE2018}
G.~Cecchini, B.~Schelter, Analytical approach to network inference:
  Investigating degree distribution, Phys. Rev. E 98.
\newblock \href {http://dx.doi.org/10.1103/PhysRevE.98.022311}
  {\path{doi:10.1103/PhysRevE.98.022311}}.

\bibitem{kenney1951}
J.~F. Kenney, E.~S. Keeping, {Mathematics of Statistics}, second edi Edition,
  Mathematics of Statistics, D. Van Nostrand Company, 1951.

\bibitem{Frank}
M.~Frank, J.~M. Buhmann, Selecting the rank of truncated svd by maximum
  approximation capacity\href {http://arxiv.org/abs/arXiv:1102.3176v3}
  {\path{arXiv:arXiv:1102.3176v3}}.

\bibitem{Gavish2014}
M.~Gavish, D.~L. Donoho, The optimal hard threshold for singular values is
  $4/\sqrt {3}$, IEEE T. Inform. Theory 60.
\newblock \href {http://dx.doi.org/10.1109/TIT.2014.2323359}
  {\path{doi:10.1109/TIT.2014.2323359}}.

\bibitem{Drossos1980}
C.~A. Drossos, A.~N. Philippou, A note on minimum distance estimates, Ann. I.
  Stat. Math. 32~(1) (1980) 121--123.
\newblock \href {http://dx.doi.org/10.1007/BF02480318}
  {\path{doi:10.1007/BF02480318}}.

\bibitem{Parr1980}
W.~C. Parr, W.~R. Schucany, Minimum distance and robust estimation, J. Am.
  Stat. Assoc. 75~(371) (1980) 616--624.
\newblock \href {http://dx.doi.org/10.1080/01621459.1980.10477522}
  {\path{doi:10.1080/01621459.1980.10477522}}.

\bibitem{Wolfowitz1952}
J.~Wolfowitz, Consistent estimators of the parameters of a linear structural
  relation, Scand. Actuar. J. 1952~(3-4) (1952) 132--151.
\newblock \href {http://dx.doi.org/10.1080/03461238.1955.10430689}
  {\path{doi:10.1080/03461238.1955.10430689}}.

\bibitem{Matusita1953}
K.~Matusita, On the estimation by the minimum distance method, Ann. I. Stat.
  Math. 5~(2) (1953) 59--65.
\newblock \href {http://dx.doi.org/10.1007/BF02949801}
  {\path{doi:10.1007/BF02949801}}.

\bibitem{wolfowitz1957}
J.~Wolfowitz, The minimum distance method, Ann. Math. Statist. 28~(1) (1957)
  75--88.
\newblock \href {http://dx.doi.org/10.1214/aoms/1177707038}
  {\path{doi:10.1214/aoms/1177707038}}.

\end{thebibliography}

\end{document}